\documentclass[prb,twocolumn,showpacs,preprintnumbers,amsmath,amssymb,citeautoscript]{revtex4-1}

\usepackage{graphicx}
\usepackage{dcolumn}
\usepackage{bm}
\usepackage{color}

\newlength{\figwidth}
\newcommand \beq {\begin{equation}} 
\newcommand \eeq {\end{equation}}

\newcommand{\ud}{\mathrm{d}}

\begin{document}

\title{Nematic and meta-nematic transitions in the iron pnictides}

\author{S. Kasahara$^{1,2}$, H.\,J. Shi$^1$, K. Hashimoto$^1$, S. Tonegawa$^1$, Y. Mizukami$^1$, T. Shibauchi$^1$,\\
K. Sugimoto$^{3,4}$, T. Fukuda$^{5,6,7}$, T. Terashima$^2$, Andriy\,H. Nevidomskyy$^8$ \& Y. Matsuda$^1$}

\affiliation{$^1$\it Department of Physics, Kyoto University, Kyoto 606-8502, Japan}

\affiliation{$^2$\it Research Center for Low Temperature and Materials Sciences, Kyoto University, Kyoto 606-8501, Japan}

\affiliation{$^3$\it Research \& Utilization Division, JASRI SPring-8, Sayo, Hyogo 679-5198, Japan}

\affiliation{$^4$\it Structural Materials Science Laboratory, RIKEN SPring-8, Sayo, Hyogo 679-5148, Japan}

\affiliation{$^5$\it Quantum Beam Science Directorate, JAEA SPring-8, Sayo, Hyogo 679-5148, Japan} 

\affiliation{$^6$\it Materials Dynamics Laboratory, RIKEN SPring-8, Sayo, Hyogo 679-5148, Japan}

\affiliation{$^7$\it JST, Transformative Research-Project on Iron Pnictides (TRIP), Chiyoda, Tokyo 102-0075, Japan}

\affiliation{$^8$\it Department of Physics and Astronomy, Rice University, 6100 Main St., Houston, TX 77005, USA}

\affiliation{$^*$\it Present address: Institute for Materials Research, Tohoku University, Sendai 980-8577, Japan}

\date{\today}

\begin{abstract}

\end{abstract}


\maketitle

{\bf Strongly interacting electrons can exhibit novel collective phases, among which the electronic nematic phases are perhaps the most surprising as they spontaneously break rotational symmetry of the underlying crystal lattice \cite{Fradkin10}. The electron nematicity has been recently observed in the iron-pnictide \cite{Zhao09,Yi-ARPES,Chuang10,Fisher-detwin,Tanatar-detwin} and cuprate \cite{Daou10,Kohsaka07,Lawler10} high-temperature superconductors. Whether such a tendency of electrons to self-organise unidirectionally has a common feature in these superconductors is, however, a highly controversial issue. In the cuprates, the nematicity has been suggested as a possible source of the pseudogap phase \cite{Daou10,Kohsaka07,Lawler10}, whilst in the iron-pnictides, it has been commonly associated with the tetragonal-to-orthorhombic structural phase transition at $T_s$. Here, we provide the first thermodynamic evidence in  BaFe$_2$(As$_{1-x}$P$_x$)$_2$ that the nematicity develops well above the structural transition and persists to the nonmagnetic superconducting regime, resulting in a new phase diagram strikingly similar to the pseudogap phase diagram in the cuprates \cite{pseudogap,Lawler10}. Our highly sensitive magnetic anisotropy measurements using microcantilever torque-magnetometry under in-plane field rotation reveal pronounced two-fold oscillations, which break the tetragonal symmetry. Combined with complementary high-resolution synchrotron X-ray and resistivity measurements, our results consistently identify two distinct temperatures---one at $T^{\ast}$, signifying a true nematic transition, and the other at $T_s (< T^{\ast})$, which we show to be not a true phase transition, but rather what we refer to as a ``meta-nematic transition'', in analogy to the well-known metamagnetic transition in the theory of magnetism. Our observation of the extended nematic phase above the superconducting dome establishes that the nematicity has primarily an electronic origin, inherent in the normal state of high-temperature superconductors. 
}

In the iron pnictides, the antiferromagnetic transition is closely intertwined with the structural phase transition from tetragonal (T) to orthorhombic (O) crystal symmetry.  Although recent experiments, including neutron scattering \cite{Zhao09}, ARPES \cite{Yi-ARPES,Shimojima}, STM \cite{Chuang10}, and transport measurements \cite{Fisher-detwin,Tanatar-detwin}, have provided evidence for electronic anisotropy, these measurements were carried out either in the low-temperature orthorhombic phase \cite{Zhao09,Chuang10,Shimojima}, where the crystal lattice structure has already broken $C_4$ symmetry, or in the tetragonal phase under uniaxial strain \cite{Yi-ARPES,Fisher-detwin,Tanatar-detwin} that also breaks this symmetry. Therefore, the question remains open whether the electronic anisotropy can exist above the structural transition without an external driving force, including under the superconducting (SC) dome. In the past, the nematic transition in the pnictides has been associated either with the orbital ordering \cite{Singh09, Lee09, Lv09, Chen09, Chen10, Lv10, andriy11}, or with the spontaneous breaking of the $Z_2$ Ising symmetry between two collinear magnetic ordering wave-vectors $\bm{Q}=(\pi,0)$ and $(0,\pi)$ \cite{Fang08, Xu08,Fernandes10,Fernandes12}.
Therefore determining the nature of the nematicity is a key to understanding the microscopic origin of the lattice and magnetic transitions, as well as its possible connection with the high-temperature (high-$T_c$) superconductivity \cite{Daou10, Lawler10}.

\begin{figure}[t]
\includegraphics[width=0.9\linewidth]{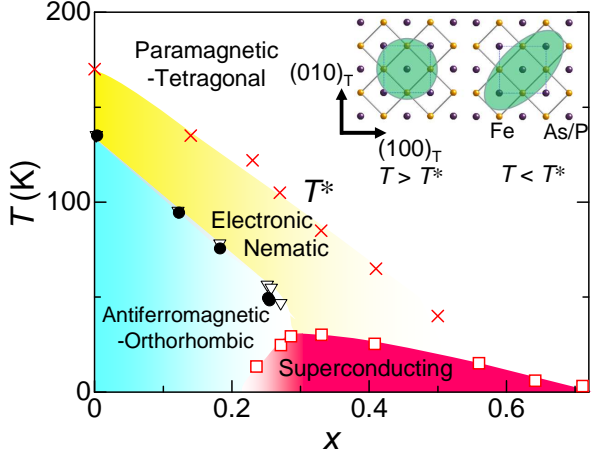}
\vskip -0.2cm 
\caption{
{\bf Doping-temperature phase diagram of BaFe$_2$(As$_{1-x}$P$_x$)$_2$.}  Solid circles, open triangles and open squares are the antiferromagnetic transition $T_N$ , T-O structural transition $T_s$, and superconducting transition $T_c$ temperatures determined by resistivity and thermal expansion measurements. Crosses indicate the nematic transition temperature $T^{\ast}$ determined by the present magnetic torque and synchrotron X-ray diffraction measurements. The insets illustrates the tetragonal FeAs/P layer.   $\chi_{ab}=0$ above $T^{\ast}$, while $\chi_{ab}\neq 0$ below $T^{\ast}$, indicating the appearance of the nematicity along [110]$_{\rm T}$ (Fe-Fe bond) direction. }
\label{Fig.PD}
\end{figure}

BaFe$_2$As$_2$-based materials are a prototypical family of iron-pnictides that play host to the superconductivity upon ion substitution. Among them, the phase diagram of the isovalent pnictogen substituted system BaFe$_2$(As$_{1-x}$P$_x$)$_2$ (ref.\:\onlinecite{Kasahara}) is shown in Fig.\:\ref{Fig.PD}. The transition to the antiferromagnetic ground state at $T_N$ always coincides or is preceded by the T-O structural transition at $T_s$. With increasing $x$, $T_N$ decreases and goes to zero continuously at $x=0.30$, indicating the presence of a quantum critical point (QCP) \cite{Nakai,Hashimoto12}.  The superconducting dome extends over a doping range $0.2<x<0.7$, with maximum $T_c=31$\,K at the QCP \cite{Kasahara}. This system is very clean and homogeneous \cite{Kasahara,Shishido,Hashimoto}, as demonstrated by the quantum oscillations observed over a wide $x$ range even in the superconducting dome \cite{Shishido}. This is important for the present work because impurities and sample inhomogeneities may otherwise wipe out the signatures of the phase transition.  Thus BaFe$_2$(As$_{1-x}$P$_x$)$_2$ appears to be the most suitable system for studying the doping evolution of the intrinsic electronic and magnetic properties in the iron-pnictides.

\begin{figure*}[bt]
\includegraphics[width=0.72\linewidth]{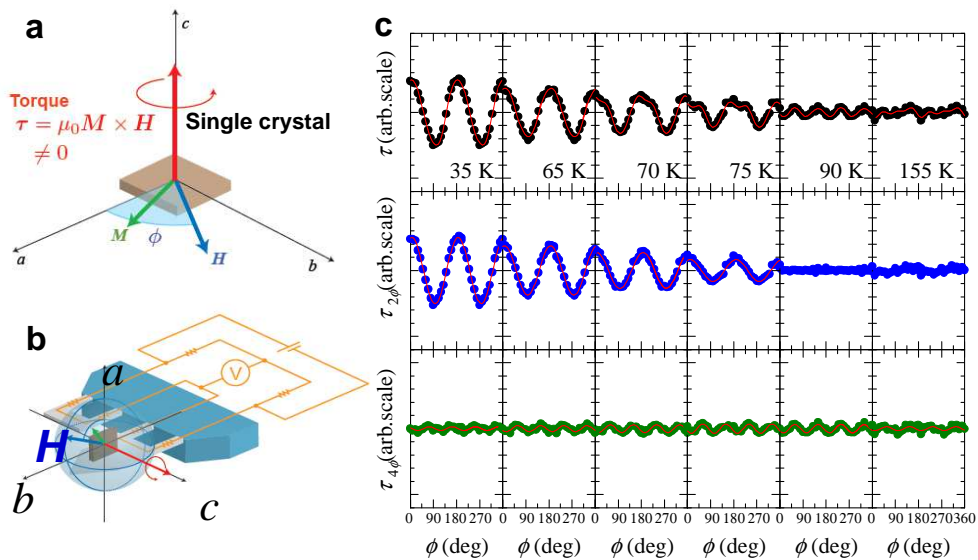}
\vskip -0.2cm 
\caption{
{\bf Torque magnetometry in the optimally doped $x=0.33$ crystal.} 
{\bf a}, {\bf b}, Schematic figures of the experimental configuration for the torque measurements under in-plane field rotation. The azimuthal angle $\phi$ is defined as the field direction from the tetragonal $a$ axis. {\bf c}, Upper panels show the raw torque data $\tau(\phi)$ at several temperatures. Middle and lower panels are the two-fold $\tau_{2\phi}$ and four-fold $\tau_{4\phi}$ components extracted from the Fourier analysis. }
\label{Fig.tau}
\end{figure*}

The magnetic torque $\bm{\tau}=\mu_0V\bm{M}\times\bm{H}$ is a thermodynamic quantity, a differential of the free energy with respect to angular displacement.  Here $V$ is the sample volume, $\bm{M}$ is the magnetization induced in the magnetic field $\bm{H}$, and $\mu_0$ is the permeability of vacuum. Torque measurement detects the magnetic anisotropy with an extremely high sensitivity \cite{Okazaki}. In particular, it provides a stringent test of nematicity. When $\bm{H}$ is rotated within the tetragonal $ab$ plane (Fig.\:\ref{Fig.tau}a, b), $\tau$ is a periodic function of double the azimuthal angle $\phi$ measured from the tetragonal $a$ axis (Fe-As/P direction, inset of Fig.\:\ref{Fig.PD}):
\begin{equation}
\tau_{2\phi}=\frac{1}{2}\mu_0H^2V[(\chi_{aa}-\chi_{bb})\sin2\phi-2\chi_{ab}\cos2\phi],
\end{equation}
where the susceptibility tensor $\chi_{ij}$ is given by $M_i =\sum_{j}\chi_{ij}H_j$.  In a system holding tetragonal symmetry, $\tau_{2\phi}$ should be zero because $\chi_{aa}=\chi_{bb}$ and $\chi_{ab}=0$.  Finite values of $\tau_{2\phi}$  appear if a new electronic or magnetic state emerges that breaks the $C_4$ tetragonal symmetry. In such a case, rotational symmetry breaking is revealed by $\chi_{aa} \neq \chi_{bb}$ and/or $\chi_{ab}\neq0$ depending on the orthorhombicity direction.  To avoid the formation of domains with different orientations in the ab-plane (`twinning'), we used very small single crystals (see Supplementary Information).

The upper panels of Fig.\:\ref{Fig.tau}c depict the temperature evolution of the torque $\tau(\phi)$ at $\mu_0H\!=\!4$\,T for optimally doped BaFe$_2$(As$_{0.67}$P$_{0.33}$)$_2$ ($T_c=30$\,K).  All torque curves are perfectly reversible with respect to the field rotation. $\tau(\phi)$ can be decomposed as $\tau(\phi)=\tau_{2\phi}+\tau_{4\phi}+\tau_{6\phi}+\cdots$, where $\tau_{2n\phi}=A_{2n\phi}\sin2n(\phi-\phi_0)$ has $2n$-fold symmetry with integer $n$.  The middle and lower panels display the two- and four-fold components obtained from the Fourier analysis. The distinct twofold oscillations appear at low temperatures, while they are absent at high temperatures. The fourfold oscillations $\tau_{4\phi}$ (and higher-order terms) are observed at all temperatures (Fig.\:\ref{Fig.tau}c, lower panels), but their amplitudes have negligible temperature dependence, indicating that they arise primarily from the nonlinear susceptibilities.   

\begin{figure*}[t]
\includegraphics[width=0.6\linewidth]{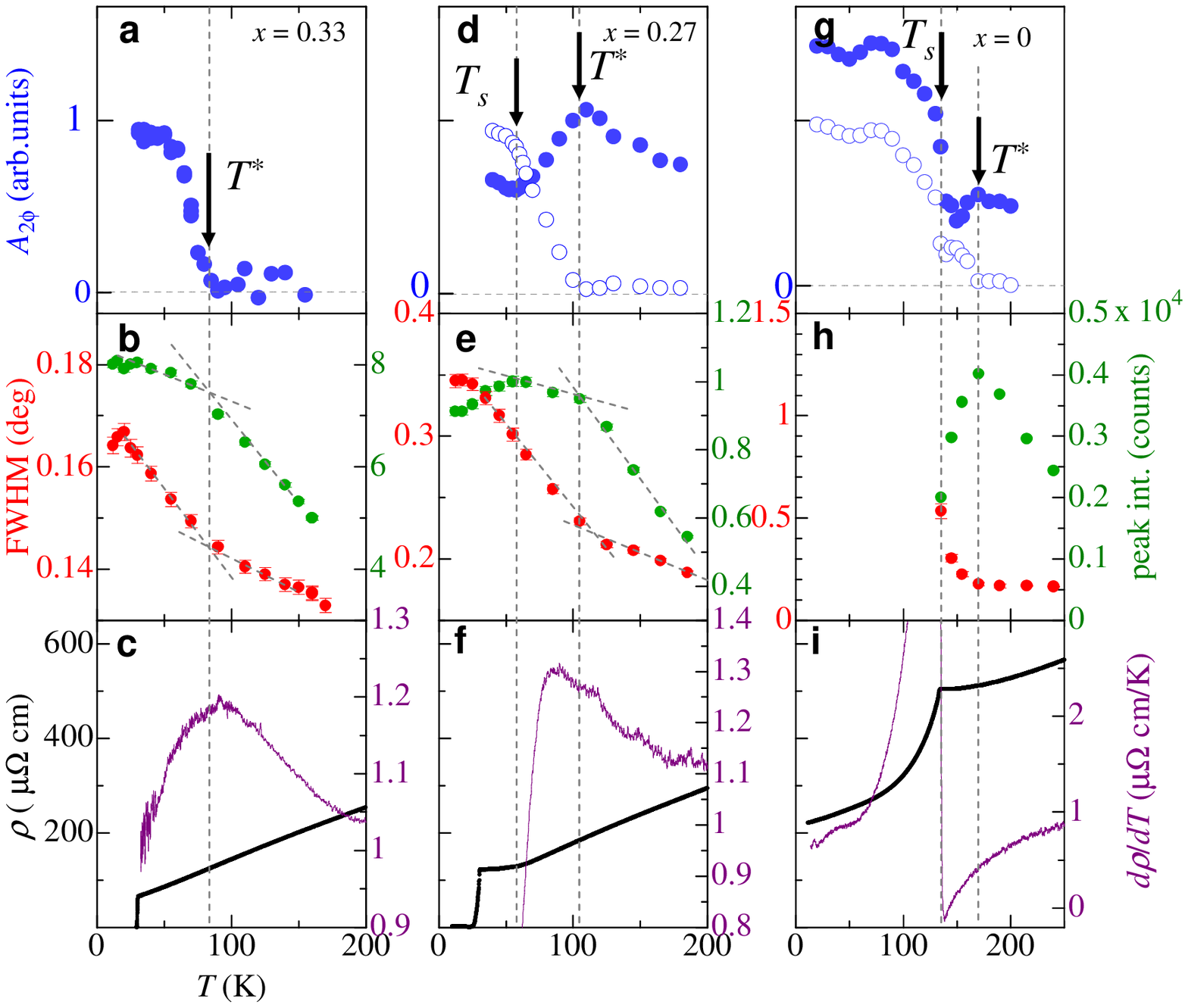}
\vskip -0.2cm 
\caption{
{\bf Temperature dependence of the two-fold oscillation amplitude $|A_{2\phi}|$ of the torque (blue circles), the FWHM (red circles) and the peak intensity (green circles) of the synchrotron X-ray Bragg reflection, and the in-plane resistivity $\rho$ (black lines).} {\bf a}-{\bf c}, Results for the crystal of $x=0.33$, {\bf d}-{\bf f}, $x=0.27$, {\bf g}-{\bf i}, $x=0$. The nematic transition temperature $T^{\ast}$ and the meta-nematic transition temperature $T_s$ are defined by arrows and vertical dashed lines. For underdoped ($x=0.27$) and parent ($x=0$) crystals, also shown are the corrected $|A_{2\phi}|$ data (open circles in {\bf d}, {\bf g}) with subtraction of the smooth background of the two-fold oscillations above $T^{\ast}$ (Fig.\:S2). The X-ray are analyzed for [10,0,0]$_{\rm T}$ ({\bf b}) and [14,0,0]$_{\rm O}$ ({\bf e}, {\bf h}) Bragg peaks and the $x=0$ data ({\bf h}) shows a clear splitting below $T_s$ (Fig.\:S3). The thin purple lines in {\bf c}, {\bf f}, {\bf i} are the temperature derivative of resistivity $d\rho/dT(T)$. 
 }
\label{Fig.combo}
\end{figure*}

As shown in Fig.\:\ref{Fig.combo}a, the amplitude of the twofold oscillation $|A_{2\phi} |$ is nearly zero at high temperatures and grows rapidly below $T^{\ast}\simeq85$\,K in the optimally doped $x=0.33$ crystals, followed by the saturation at lower temperatures. The anomaly at $T^{\ast}$ can also be seen by the synchrotron X-ray diffraction and in-plane resistivity measurements. Figure\:\ref{Fig.combo}b depicts the temperature dependencies of the full width at half maximum (FWHM) and intensity of the Bragg peak below 200\,K. Both the FWHM and the peak intensity change their slope in the vicinity of  $T^{\ast}$.  Moreover, the temperature derivative of the resistivity $d\rho/dT$ exhibits a peak at $\sim T^{\ast}$ (Fig.\:\ref{Fig.combo}c).  Here we stress that sharp synchrotron X-ray diffraction peaks, NMR spectra which show no indication of magnetic ordering \cite{Nakai} and purely paramagnetic response above $T_c$ (Fig.\:S1), all indicate that the sample inhomogeneity is highly unlikely to be an origin of the anomaly at $T^{\ast}$.

\begin{figure}[t]
\includegraphics[width=0.75\linewidth]{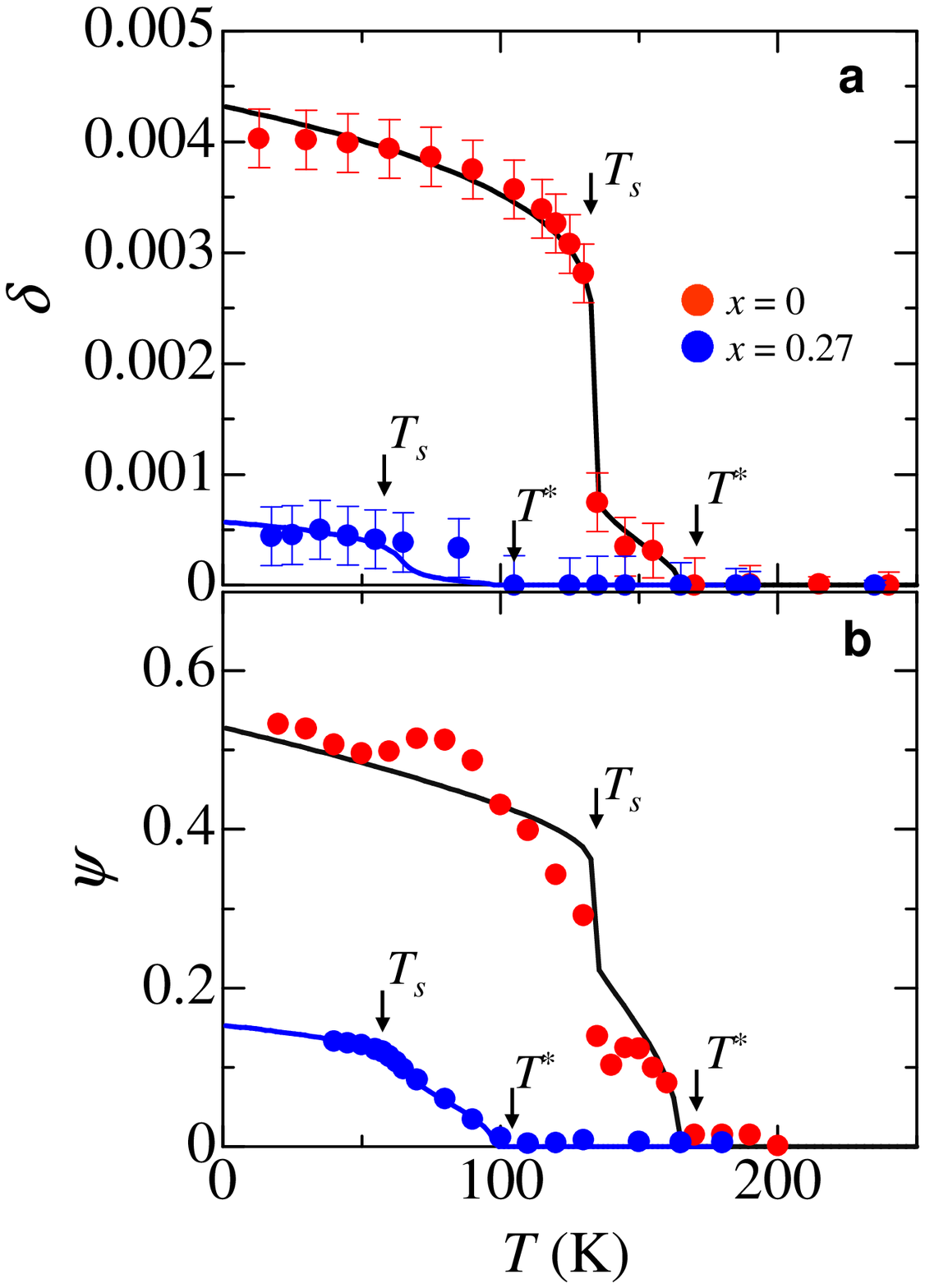}
\vskip -0.25cm 
\caption{{\bf The fits (solid lines) using the Landau free energy expansion Eq.\:(\ref{landau}) to experimental data (solid circles)}. {\bf a}, Lattice distortion $\delta=\frac{a-b}{a+b}$ and {\bf b}, the nematic order parameter $\psi$ which is proportional to measured $A_{2\phi}$ component of the torque are fitted using the same set of parameters: 
for the parent compound ($x=0$, red symbols): $u=3.277,\, v=4.078,\, w=4.523,\, g= 1.905$, $T_s^{(0)}=52$\,K, $T_p^{(0)}=117$\,K; for the $x=0.27$ compound (blue circles): $u=0.988,\,  v=5.728,\,  w=73.084,\, g=1.299$, $T_s^{(0)}=48$\,K,  $T_p^{(0)}=70$\,K.}
\label{Fig.landau}
\end{figure}

The above results clearly indicate that the tetragonal $C_4$ symmetry, which is preserved at high temperatures, is broken below $T^{\ast}$, demonstrating the formation of the electronic nematic phase at $T^{\ast}$. The twofold oscillation below $T^{\ast}$  follows the functional form,  $\tau_{2\phi}=A_{2\phi}\cos2\phi$, meaning that $\chi_{aa}=\chi_{bb}$ and $\chi_{ab}\neq 0$, which indicates the nematicity along the tetragonal [110]-direction, i.e. Fe-Fe bond direction (Fig.\:\ref{Fig.PD}, inset). The X-ray FWHM at $T<T^{\ast}$ (Fig.\:\ref{Fig.combo}b, red circles) grows slightly larger than the linear extrapolation from above $T^{\ast}$ (dashed line), and is accompanied by the suppression of the X-ray peak intensity (green circles). This indicates a broadening of the Bragg peak below $T^{\ast}$, implying that the nematicity to some extent couples to the orthorhombic lattice distortion as discussed later.

Figures \ref{Fig.combo}d--f show the temperature dependence of $|A_{2\phi}|$, X-ray peaks and resistivity, respectively, of the underdoped material ($x=0.27$). It undergoes the \text{T-O} transition at $T_s\sim 60$\,K, and the magnetic transition well below $T_c$.  $|A_{2\phi}|$ is finite even at 200\,K, and initially increases with decreasing temperature, exhibiting a cusp-like peak at $T^{\ast}\simeq 105$\,K, which we attribute to the nematic transition temperature for the following reasons. 
Analysis of more than five samples showed that the anomaly at 105\,K (and also at $T_s$) is well reproduced, while the torque curves above 105\,K show some degree of variation depending on the sample (Fig.\:S2). In fact, when the temperature dependent part above 105\,K is subtracted, $|A_{2\phi}|$ (Fig.\:\ref{Fig.combo}d, open circles) exhibits very similar temperature dependence to that in $x=0.33$ compound (Fig.\:\ref{Fig.combo}a). In addition, it is apparent that the FWHM and intensity of X-ray peak (Fig.\:\ref{Fig.combo}e) change their slope at $T \simeq 105$\,K, quite similarly to those of the optimally doped compound at $T^{\ast}$ (Fig.\:\ref{Fig.combo}b). Moreover, the hump structure can be seen in the $d\rho/dT$ at around the same temperature (Fig.\:\ref{Fig.combo}f).

At $T_s\sim 60$\,K, $|A_{2\phi}|$ reaches a minimum and increases again at lower temperatures, indicating that $|A_{2\phi}|$ is sensitive to the T-O structural transition. At temperatures above $T^{\ast}$, the origin of the non-zero twofold signal is not clear. It may be due to the presence of local impurities or dislocations.  In fact, local orbital ordering with $C_2$ symmetry around the impurity sites has been proposed \cite{Kontani}. 

Figures\:\ref{Fig.combo}g--i show the results for the parent compound $x=0$, in which the structural transition and magnetic order occurs at the same temperature  $T_s=T_N$. Similar to underdoped compound,  $|A_{2\phi}|$ is finite in the high temperature regime.  As the temperature is lowered, $|A_{2\phi}|$ increases gradually and then decreases with a cusp-like peak at $\sim 170$\,K.  At the same temperature, the synchrotron X-ray diffraction exhibits anomalies of the FWHM and peak intensity (Fig.\:\ref{Fig.combo}h). Thus, by the same reasoning as for the underdoped compound, the cusp-like behaviour of $|A_{2\phi}|(T)$ is attributed to the nematic transition; $T^{\ast}\simeq170$\,K. Behaviour of $|A_{2\phi}|(T)$ at $T_s$ is also similar to underdoped compound but is more pronounced; $|A_{2\phi}|$ is strikingly enhanced below $T_s$, which is attributed to the larger distortion associated with the T-O transition. The resistivity shows a rather flat depenendece and no apparent anomaly is observed at $T^{\ast}$.   Both  $|A_{2\phi}|$ and $d\rho/dT$ exibit anomalies at around 50\,K, which may indicate the presence of the magnetic or charge ordering deep inside the antiferromagnetic phase.

We emphasize that in contrast to previous experiments, the present measurements were performed without applying external pressure or uniaxial stress. Thus our results provide thermodynamic evidence for the spontaneous formation of the electronic nematic phase below $T^{\ast}$, well above the previously reported T-O structural transition temperature $T_s$ in this system. 
Moreover, the doping dependence of the nematic transition $T^{\ast}(x)$ displayed in Fig.\:\ref{Fig.PD} indicates that nematicity develops independently of $T_s(x)$ 
and persists over a wide range of doping covering the non-magnetic superconducting regime. 

Clearly, there cannot be two nematic phase transitions at both $T_s$ and $T^{\ast}$, because the $C_4$ rotational symmetry can only be broken once. The temperature $T^{\ast}(>T_s)$ marks the onset of the true phase transition, accompanied by the nematic two-fold torque component $A_{2\phi}\neq 0$. Then what happens to the structural transition at $T_s$? This question can be answered straightforwardly if one considers the Landau free energy expansion in terms of two parameters, the lattice distortion $\delta$ and the nematic parameter $\psi\propto A_{2\phi}$, which  can be written as follows:
\begin{equation}
F[\delta,\psi] = \left[ t_s\delta^2 - u\delta^4 + v\delta^6 \right] 
+ \left[ t_p\psi^2 + w\psi^4 +\mathcal{O}(\psi^6)\right] - g\,\psi\cdot\delta, 
\label{landau}
\end{equation}
with the terms in the first square bracket describing the first-order structural phase transition and the second bracket responsible for the (second-order)  nematic phase transition. The temperature-dependent coefficients $t_s=(T-T_s^{(0)})/T_s^{(0)}$ and $t_p=(T-T_p^{(0)})/T_p^{(0)}$ were chosen such that in the absence of the coupling between the two order parameters, the structural transition occurs  at lower temperature $T_s^{(0) }(< T_p^{(0)})$. Thanks to the linear coupling between the two order parameters, as expressed in the last term, both $\psi$ and $\delta$ develop non-zero values below the transition temperature $T^{\ast}$ (Figs.\:\ref{Fig.landau}a,b). 
 On the other hand $T_s$ ceases to be a true phase transition, since the $C_4$ symmetry is broken on either side of $T_s$, and lattice distortion $\delta$ is non-zero over the entire temperature range (Fig.\:\ref{Fig.landau}a). Instead, both $\delta$ and $\psi$ undergo a finite jump at $T_s$, as illustrated in Fig.\:\ref{Fig.landau}. We call this a \emph{meta-nematic} transition, in analogy to the meta-magnetic transition in the theory of magnetism, where the magnetization undergoes a jump as a function of temperature or applied magnetic field, but remains non-zero on both sides of the transition. The analysis of the free energy shows that it exhibits a maximum at $\psi=0$ and a single minimum at finite $\psi$, as in the $2^\text{nd}$-order Landau phase transition (Fig.\:S4). Note that because of the coupling between the order parameters, both $T_s$ and $T^{\ast}$ are renormalized compared to their initial values $T_s^{(0)}$ and $T_p^{(0)}$ (see Supplementary Information for details).

To quantify the lattice distortion $\delta$ experimentally, we have analysed the X-ray data by using two-peak fitting (Fig.\:S3), which reveals that the data in the region $T_s<T<T^{\ast}$ can be fitted with very small but finite $\delta$. The obtained results of $\delta(T)$ can be reasonably reproduced within the framework of Eq.\:(\ref{landau}), and the same set of Landau parameters also fits well the temperature dependence of  $\psi\propto A_{2\phi}$ (Fig.\:\ref{Fig.landau}).
We have thus established that the true thermodynamic transition occurs at $T=T^{\ast}$, and is accompanied by the development of the non-zero values of both the nematic order parameter $\psi$ and the lattice distortion $\delta$. 
We note that similarly small but non-zero values of $\delta$ have been recently reported in powder diffraction measurements of SmFeA(O$_{1-x}$F$_x$) \cite{Sm-1111}.

Note that this explanation is very generic and does not depend on the precise microscopic nature of the nematic transition, be it caused by $Z_2$ spin-nematic ordering \cite{Fang08,Xu08,Fernandes10,Fernandes12}, or by orbital ordering \cite{Singh09,Lee09,Lv09,Chen09,Chen10,Lv10,andriy11}. 
In the spin-nematic approach, the nematic instability is driven by thermal spin fluctuations above the SDW ordered phase \cite{Fernandes12}, which would apply to the $x \lesssim 0.30$ regime in BaFe$_2$(As$_{1-x}$P$_x$)$_2$. Such fluctuations have been detected by NMR \cite{Nakai}. However, the fact that the nematic transition at $T^*$ occurs even for superconducting samples well above optimal doping, and far away from the SDW phase, provides a strong indication that the nematic transition is unlikely to be associated with thermal fluctuations above long-range magnetic order. On the other hand, in the case of orbital ordering, 
the nematic transition naturally occurs as a result of polarisation between the Fe $d_{xz}$ and $d_{yz}$ orbitals, $\psi \propto (n_{xz} - n_{yz})$, as supported by the recent ARPES \cite{Yi-ARPES} and quadrupolar resonance measurements \cite{NQR-1111}.

A large anisotropy in resistivity $(\rho_a - \rho_b)/(\rho_a + \rho_b)$ has been recently observed under the uniaxial stress in Ba(Fe$_{1-x}$Co$_x$)$_2$As$_2$ at temperatures higher than $T_s$ and even above the superconducting dome. In the light of the present results, this anisotropy can be explained as associated with the nematic transition at $T^{\ast}$, rather than with a proximity to the structural phase transition. The application of a uniaxial stress will enhance the nematicity and is likely to shift $T^{\ast}$ to a higher temperature.

There is growing body of evidence that entanglement of the spin and orbital degrees of freedom leads to emergent novel electronic phases in the iron pnictides. The present temperature-doping phase diagram bears striking resemblance to that of high-$T_c$ cuprates, in that the suppression of the antiferromagnetic ground state leads to the emergence of high-$T_c$ superconductivity
and electron nematic instability occurs well above the magnetic and superconducting transitions. Recent infrared studies of charge dynamics report the formation of a pseudogap in the excitation spectrum of optimally doped BaFe$_2$(As$_{1-x}$P$_x$)$_2$ below $\sim100$\,K (Moon, S.\,J. {\it et al.}, unpublished results). It is therefore likely that the nematic transition is related to the pseudogap formation, similar to the underdoped cuprates. These electronic properties may capture a universal feature essential for the occurrence of high-$T_c$ superconductivity. \\

{\bf Acknowledgements.} We thank fruitful discussion with A.\,V. Chubukov, R.\,M. Fernandes, I. Fischer, H. Ikeda, H. Kontani, and R. Okazaki. 
This research has been supported through Grant-in-Aid for the Global COE program ``The Next Generation of Physics, Spun from  Universality and Emergence" from MEXT of Japan, and KAKENHI from JSPS.   A.H.N. and Y.M. acknowledge the hospitality of the Aspen Center for Physics.\\

\newpage

\renewcommand{\refname}{Supplementary References}
\renewcommand{\thesection}{\Roman{section}}
\renewcommand{\thesubsection}{\Alph{subsection}}

\setcounter{figure}{0}
\renewcommand{\figurename}{Figure S$\!\!$}

\newpage
\newpage
\begin{center}
{\large\bf Supplementary Information}
\end{center}

\section{Methods}

High-quality single crystals are grown by the self-flux method \cite{Kasahara}. To measure the magnetic anisotropy accurately, the microcantilever torque magnetometry is used (Fig.\:2b). To avoid the torque oscillation arising from the misalignment, the magnetic field $\bm{H}$ is precisely applied in the $ab$ plane within an error less than 0.1$^\circ$ by controlling two superconducting magnets and the rotating stage \cite{Okazaki}. 

In a nematic state, the domain formation with different preferred directions in the $ab$ plane may occur as a consequence of the degeneracy in tetragonal crystal structure (``twinning'').  If the domain size is much smaller than the crystal size, the amplitude of $\tau_{2\phi}$ would be significantly diminished due to the cancellation of the two-fold oscillations with opposite sign arising from different domains.  To avoid such an effect, we used very small single crystals with typical size $\sim70\times70\times30\,\mu$m$^3$. 

In order to examine the detailed structural change of the underlying crystal lattice, high-resolution structure analysis by using the synchrotron X-ray crystallography techniques were performed on the same crystals used in the torque measurements and the high angle diffraction spots at tetragonal [10,0,0]$_{\rm T}$ or orthorhombic [14,0,0]$_{\rm O}$ were analysed. 
The sample temperature was controlled by an open flow cryocooler, whose temperature reading may have a difference from the actual sample temperature by up to $\sim2$\,K, but this is sufficiently small for the purpose of discussion of the phase diagram in Fig.\:\ref{Fig.PD}.

\section{Out-of-plane magnetic response}

All of the single crystals used in the present study exhibit purely paramagnetic response above $T_N$ and $T_c$, which was carefully checked by measuring the angular variation of the torque in $\bm{H}$ rotating within the $ac$ plane. This excludes the possibility of extraneous magnetic impurities in the samples.

Figure S1a depicts the torque measured in field {\boldmath $H$} rotated within the $ac$ plane for underdoped ($x=0.27$) single crystal at $\mu_0H$=4~T.  The curves in this geometry are perfectly sinusoidal, well fitted with $\tau(T,H,\theta)=A_{2\theta}(T,H)\sin2\theta$, where $A_{2\theta}$ is the amplitude and $\theta$ is the polar angle.    The hysteresis component is less than 0.01\% of the total torque, indicating no detectable ferromagnetic impurities.   In this geometry, the difference $\Delta\chi_{ca}$ between the $c$ axis and in-plane susceptibility yields a two-fold oscillation from $\tau_{2\theta}(\theta,T,H)$ with respect to $\theta$ rotation
\begin{equation}
\tau_{2\theta}=\frac{1}{2}\mu_0H^2V\Delta\chi_{ca}\sin 2\theta. \tag{S-1}
\end{equation}

The $H$-linear dependence of $A_{2\theta}(H,T)/\mu_0 H$ has null $y$ intesect (Fig.\:S1b), which indicates a field-independent magnetic susceptibility,  that is a purely paramagnetic response.     This also reinforces the absence of ferromagnetic impurities.    As the temperature is lowered, $\Delta\chi_{ca}=\chi_{cc}-\chi_{aa}$ increases gradually and then decreases with a peak at the T-O structural transition $T_s$.  

\begin{figure*}[t!]
\vspace{-2mm}
\includegraphics[width=150mm]{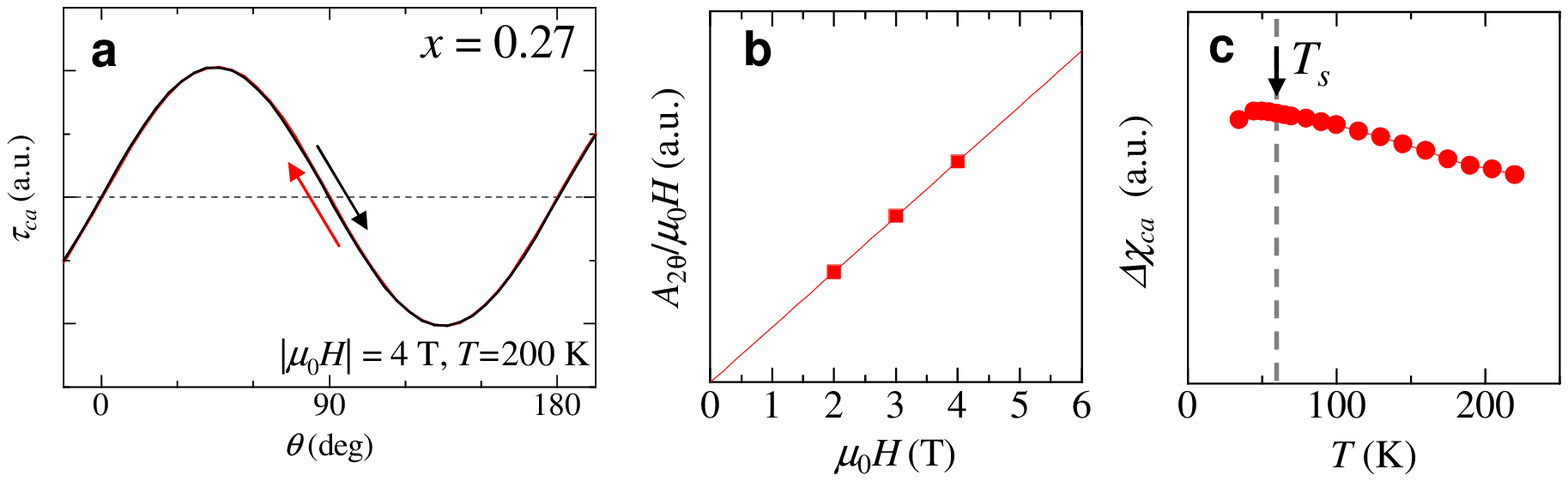}
\caption{(a)The torque curve $\tau_{ac}(\theta)$ in {\boldmath $H$} rotating within the $ac$ plane.  (b)$A_{2\theta}/\mu_0H$ plotted as a function of $\mu_0H$.  (c) Temperature dependence of  $\Delta\chi_{ca}$.   }
\end{figure*}

\section{In-plane magnetic response and sample dependence}

Figure\:S2 depicts the sample dependence of $A_{2\phi}$ for underdoped ($x=0.27$) compound.   In both crystals, anomaly at $T^{\ast}$ is clearly seen.   Solid symbols indicate the data, in which the temperature dependent $A_{2\phi}$ at $T>T^{\ast}$ is subtracted by assuming linear-$T$ dependence. After subtracting the high-temperature $A_{2\phi}$, which may be due to the local impurities or dislocations, $A_{2\phi}(T)$ shows nearly identical temperature dependence. Such a dependence is consistent with the raw data of $x=0.33$, in which the high-temperature term is absent. 


\begin{figure}[t]
\begin{center}\leavevmode
\includegraphics[width=0.65\linewidth]{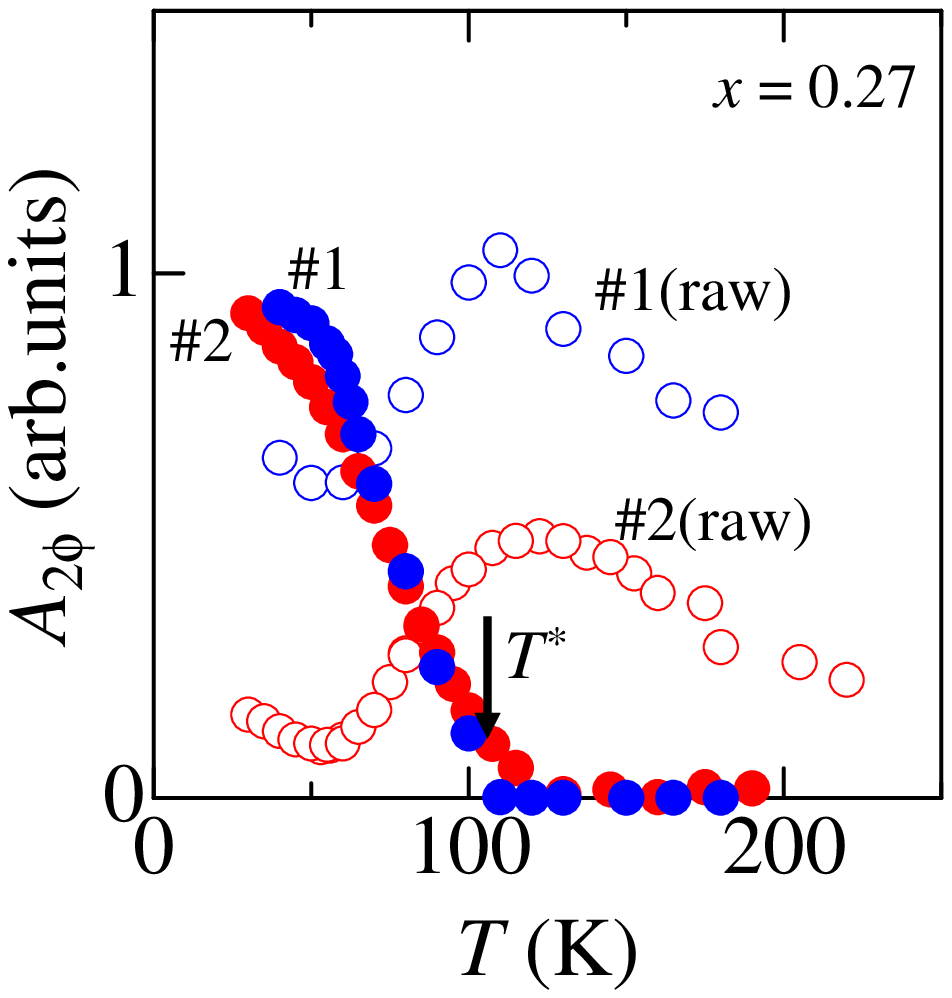}
\caption{Temperature dependence of $A_{2\phi}$ for two samples of $x=0.27$. The open symbols are the raw data. At high temperatures, the $A_{2\phi}(T)$ data are fitted with the $T$-linear dependence $A_{2\phi}^{\rm HT}(T)$ and from each torque curve $\tau(\phi)$ the two-fold oscillation term with $A_{2\phi}^{\rm HT}(T)$ amplitude is subtracted to extract the reanalysed $A_{2\phi}(T)$ (solid symbols).}
\end{center}
\end{figure}

%

\section{Synchrotron X-ray diffraction and crystal structure analysis}

Synchrotron X-ray diffraction were  perfomed on BL02B1 at SPring-8 on the same crystals used in the torque measurements. The photon energy of the incident X-ray was tuned at 17.7 keV. 
A large cylindrical imaging plate and Rapid-auto program (Rigaku Corp.) was used to obtain diffraction data \cite{Sugimoto}. 
The sample temperature was controlled by an open flow cryocooler, whose temperature reading may have a difference from the actual sample temperature by up to $\sim2$\,K, but this is sufficiently small for the purpose of discussion of the phase diagram in Fig.\:1 in the main text.

Figure\:S3a depicts the temperature dependence of diffraction angle 2$\theta$ at orthorhombic (14,0,0)$_{\rm O}$ for parent ($x=0$) and underdoped ($x=0.27$) crystals.  The Bragg peak is analyzed by the two Gaussian curves (Fig.\:S3c). Below $T^\ast$, the data can be fitted with two peaks, from which the orthorhobic distortion $\delta=\frac{a-b}{a+b}$ is estimated (Fig.\:S3b). 

\begin{figure*}[t!]
\vspace{-2mm}
\begin{center}\leavevmode
\includegraphics[width=0.7\linewidth]{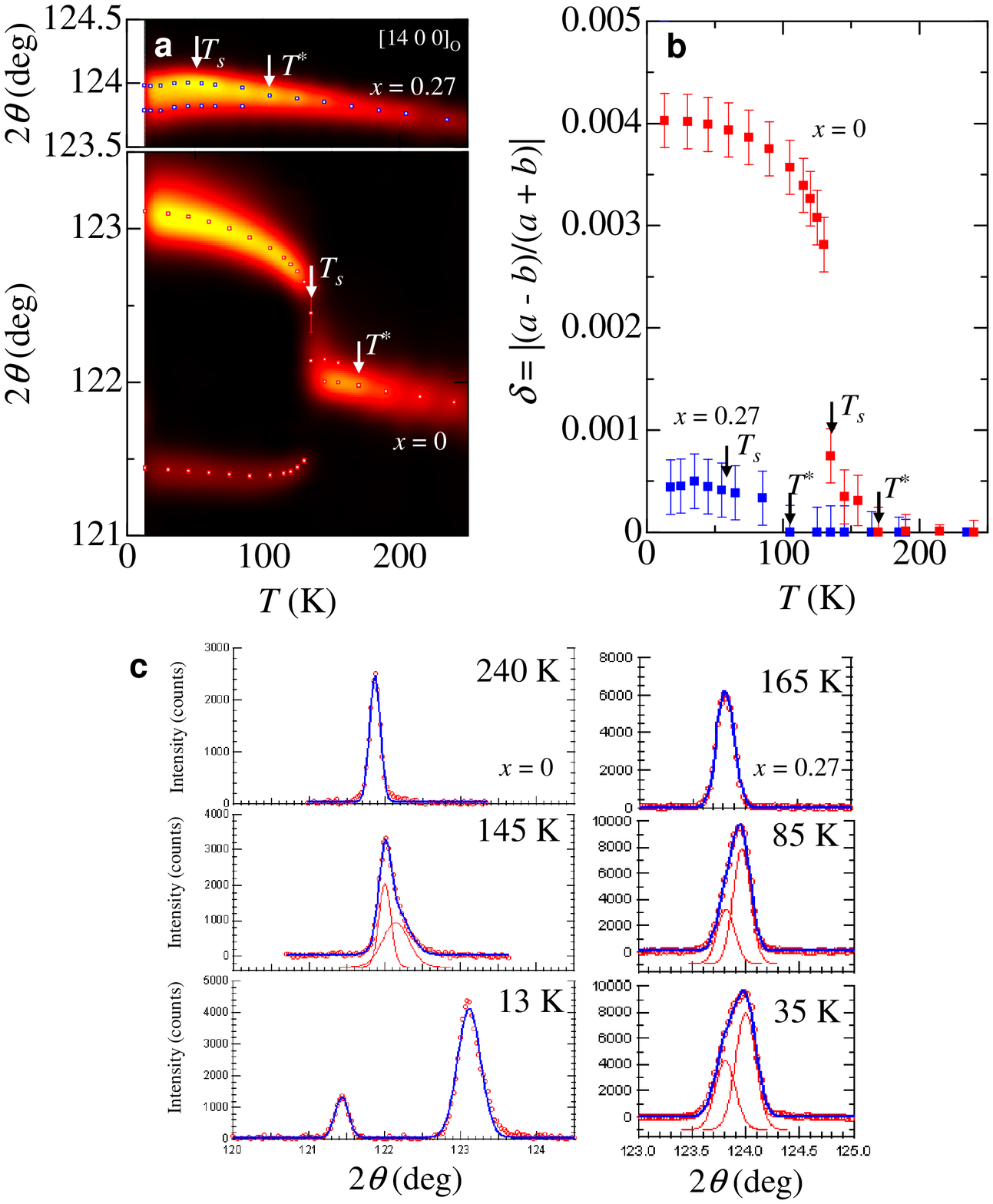}
\caption{Temperature dependence of $2\theta$ (a) and $\delta=|\frac{a-b}{a+b}|$ (b) for $x=0$ and 0.27. The color in (a) indicates the Bragg peak intensity. (c) Typical diffraction data (red circles) near the [14\,0\,0]$_{\rm O}$ Bragg peak are fitted (blue lines) with two overlapping peaks (red curves) for $T>T^\ast$ (upper panels), $T^\ast>T>T_s$ (middle panels), and $T_s>T$ (lower panels).}
\end{center}
\end{figure*}

The increase of the FWHM with decreasing $T$ below 200\,K down to $T^{\ast}$ (Fig.\:3b, e, h) is unusual because this temperature dependence is opposite to what is expected from the thermal vibration of atoms. This may indicate the presence of rather strong nematic fluctuations well above $T^{\ast}$. We also note that the FWHM is supressed when the SC transition sets in.  This implies the coupling between the nematic order and superconductivity. The peak of $d\rho/dT$ at $T^{\ast}$ (Fig.\:3c, f) suggests that the formation of the nematicity affects the electronic transport properties, possibly due to scattering off the nematic domain boundaries.


\section{Landau free energy analysis}

The Landau free energy is expressed in Eq.\:(2) in the main text in terms of the nematic order parameter $\psi$ and  the structural orthorhombic distortion  $\delta=(a-b)/(a+b)$ as follows:
\begin{equation*}
F[\delta,\psi] = \left[ t_s\delta^2 - u\delta^4 + v\delta^6 \right] 
+ \left[ t_p\psi^2 + w\psi^4 +\mathcal{O}(\psi^6)\right] - g\,\psi\cdot\delta.
\end{equation*}
The quadratic coefficients are proportional to the reduced temperatures of the respective transitions:
\begin{equation}
t_s = \frac{T-T_s^{(0)}}{T_s^{(0)}};\qquad t_p = \frac{T-T_p^{(0)}}{T_p^{(0)}}, \tag{S-2}
\end{equation}
where $T_s^{(0)}$ and $T_p^{(0)}$ have a meaning of the structural and nematic transition temperatures, respectively, in the absense of the coupling between the two order parameters ($g=0$). These are the ``bare'' transition temperatures, which will become renormalized due to the electron-lattice coupling, as shown below.

The saddle-point solution of the Landau free energy can be found from the following two equations:
\begin{align}
g\,\psi  &\,=\, 2t_s \delta - 4u\,\delta^3 + 6v\,\delta^5 \tag{S-3a} \\
g\,\delta &\,=\,  2t_p \psi + 4w\, \psi^3          \tag{S-3b}
\end{align}
The last equation can be thought of as expressing the structural distortion in terms of the nematic order parameter $\psi$: $\delta = \delta(\psi)$. The free energy can then be expressed in terms of $\psi$ alone: $F(\psi)=F[\psi,\delta(\psi)]$, and is plotted in Fig.\:S4a for two different temperatures below (above) $T_s$, shown in solid (dashed) line. In both cases, the free energy has a maximum at $\psi=0$ and a single minimum at finite $\psi$, so that it behaves as in the $2^\text{nd}$-order Landau phase transition. This is a general result, provided the quartic coefficient $w$ is larger than $|u|$ (in the opposite limit, $|u|\gg w$, the coupling between the two order parameters can render the nematic transition to be first order, but this does not alter the rest of our conclusions).

\begin{figure*}[t!b]
\includegraphics[width=0.7\linewidth]{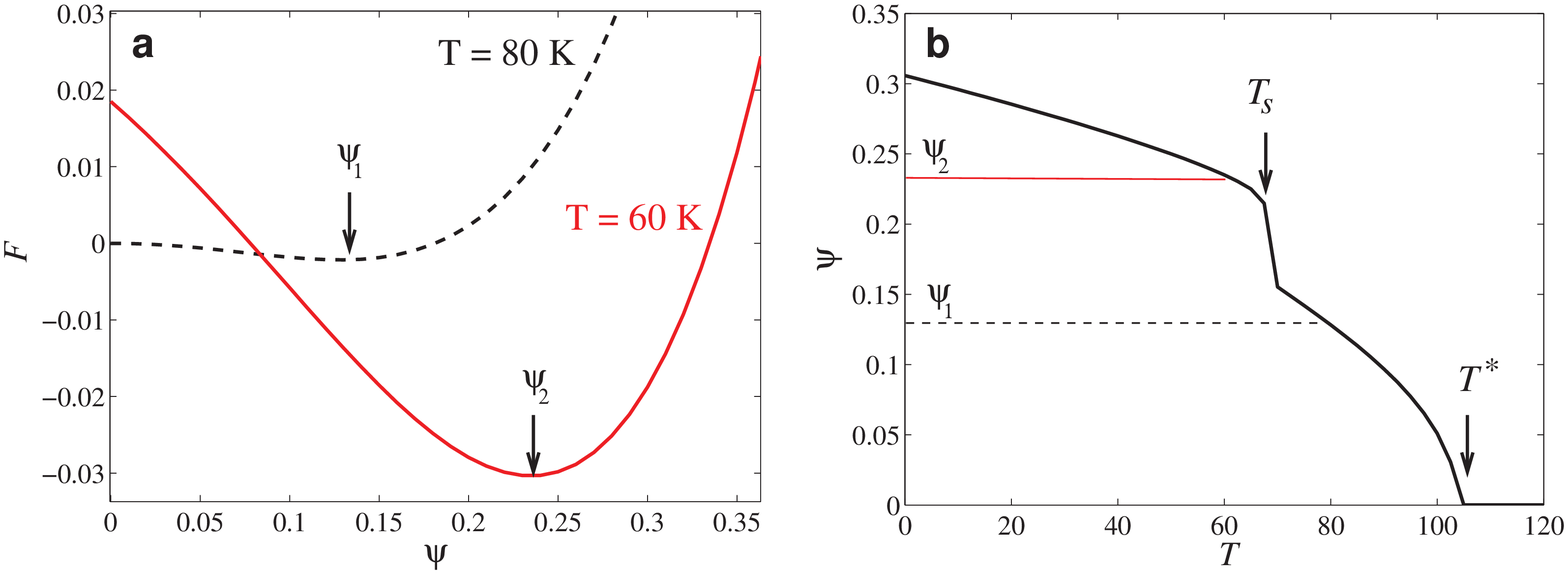}
\caption{ Landau free energy analysis, Eq.\:(2) in the main text, with parameters $u\!=\!4, v\!=\!10, w\!=\!8, g\!=\!0.5$ and $t_s(T), t_p(T)$ expressed  as in Eq.\:(S-2)  via the unrenormalized transition temperatures  $T_s^{(0)}=40$\,K, $T_p^{(0)}=80$\,K. {\bf a}, free energy $F(\psi)$ as a function of the nematic order parameter $\psi$ for two temperatures, $T=60$\,K and $80$\,K, on either side of the meta-nematic transition at $T_s\approx 70$\,K. {\bf b}, the saddle-point solution of Eqs.\:(S-3) for the nematic order parameter $\psi (T)$. 
}
\label{Fig.Landau}
\end{figure*}

We observe that the saddle-point solution, obtained by solving the equation $\ud F/\ud \psi=0$, changes discontinuously when temperature goes through $T_s$, as illustrated in Fig.\:S4b. Consequently, structural distortion $\delta$ also sustains a jump at $T_s$, marking the onset of the \emph{meta-nematic transition}, as explained in the main text. This is not a true phase transition because both order parameters $\psi$ and $\delta$ have finite values on either side of $T_s$, as illustrated in Fig.\:4a, b in the main text. 

The true phase transition occurs at a higher temperature $T^*$, which can be obtained from the saddle-point equations by noting that sufficiently close to $T^*$, the $\psi^3$ term can be neglected in Eq.\:(S-3b), resulting in
\begin{equation*}
\delta \approx \frac{g}{2t_s}\psi, \quad \text{for } T_s<T<T^*.  \tag{S-4}
\end{equation*}
It then follows from Eq.\:(S-2a) that 
\begin{equation*}
\psi^2 \approxeq \frac{1}{4w}\left(-2t_p + \frac{g^2}{2t_S}\right).  \tag{S-5}
\end{equation*}
This last equation has a solution for $T<T^*$, and denoting $t_p$ in terms of the reduced temperature $t_p=(T-T_p^{(0)})/T_p^{(0)}$ as in Eq.\:(S-2), we can express $T^*$ as follows:
\begin{equation*}
T^* \approxeq T_p^{(0)} \left( 1 + \frac{g^2}{4t_s}\right). \tag{S-6}
\end{equation*}
$T_p^{(0)}$ has a meaning of the nematic transition temperature in the absence of coupling to the lattice ($g=0$). Because of this coupling, the transition is shifted to a higher temperature $T^* > T_p^{(0)}$.

\begin{figure}[t!b]
\includegraphics[width=0.7\linewidth]{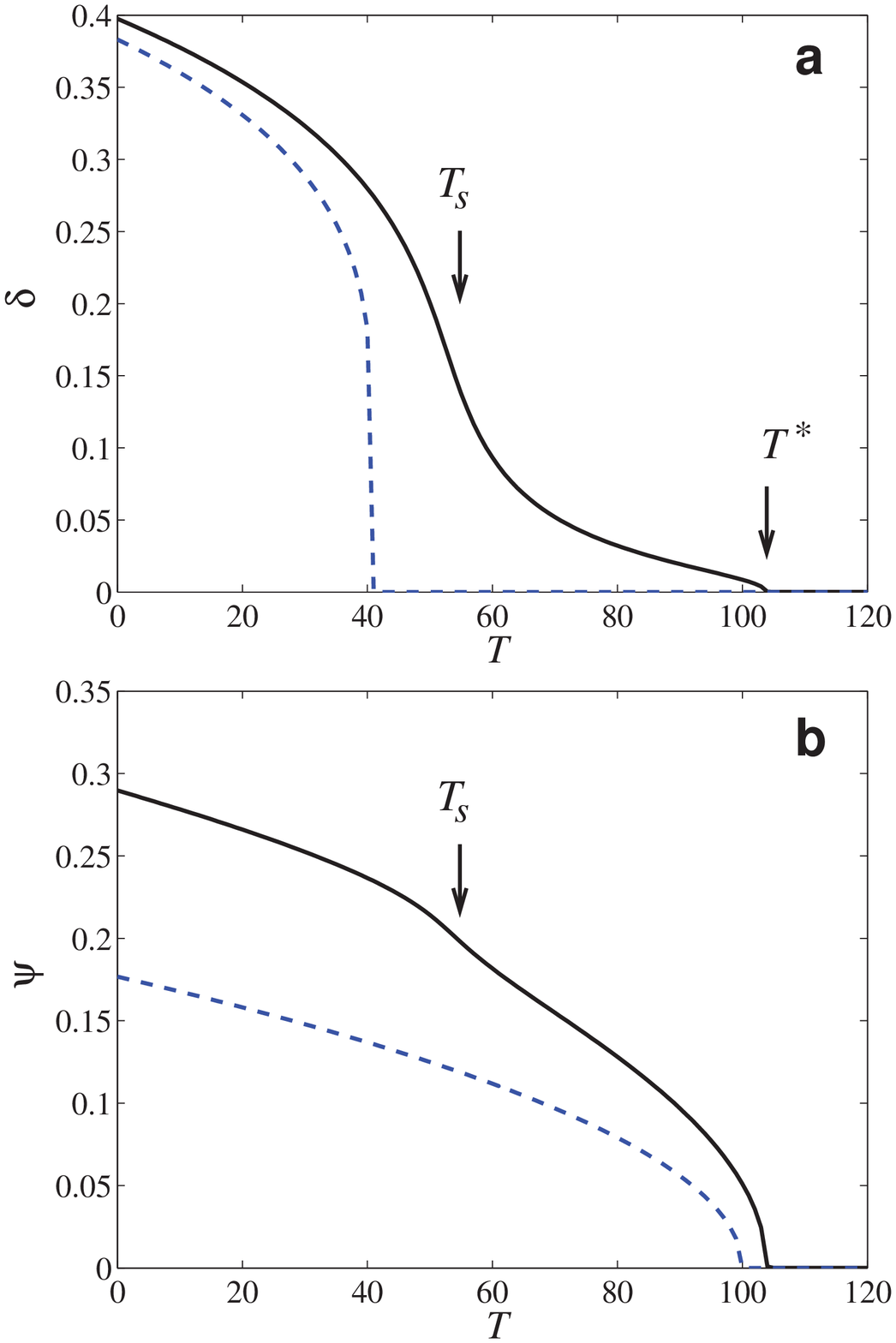}
\caption{Landau free energy analysis with the same parameters as in Fig.\:S4, except  $u=1$ and $v=20$. Plotted with solid lines are the saddle-point solutions of Eqs.\:(S-3) for {\bf a}, the lattice distortion $\delta(T)$ and {\bf b}, the nematic order parameter $\psi(T)$. The dashed lines denote the respective solutions in the absence of a coupling between the  order parameters ($g=0$ in Eq.~2).}
\label{Fig.Landau2}
\end{figure}

Note that in the absence of coupling to the lattice ($g=0$) the second-order electronic nematic phase transition at $T_c^{(0)}$ would have been completely decoupled from the first-order structural transition at $T_s < T_c^{(0)}$, as illustrated with the dashed lines in Fig.\:S5. However since both the nematic and orthorhombic order parameters break the $C_4$ rotational symmetry of the crystal, it is natural to expect a non-vanishing coupling between the two. As a result, the orthorhombic distortion $\delta$ develops a non-zero value below $T^*$, rendering the transition at $T_s$ a \emph{meta-nematic}, instead of a true first-order structural transition. This is corroborated by our X-ray diffraction studies, which find that the 
Bragg peak starts broadening when temperature is lowered below $T^*$, illustrated in Fig.\:3b, e, h in the main text.

The sharpness of the  meta-nematic transition at $T_s$ depends on the strength of the coupling $g$ between the order parameters relative to the typical jump in $\delta$ at $T_s$, $\Delta\delta$. To illustrate the point, we plot in Fig.\:S5  the saddle-point solutions of the Landau free energy using a different set of expansion parameters. 
As can be seen in Fig.\:S5, the meta-nematic transition becomes much smoother and less pronounced than in Fig.\:4 in the main text, matching qualitatively the experimental data in the $x=0.27$ underdoped sample, see Fig.\:4 in the main text.

The above arguments are very general and do not depend on the microscopic origin of the electronic nematic order parameter $\psi$. It has been argued~\cite{Singh09, Lee09, Lv09, Chen09, Chen10, Lv10, andriy11} 
that the nematicity may be driven by orbital physics, due to spontaneous imbalance between the population of the Fe $d_{xz}$ and $d_{yz}$ orbitals, in which case $\psi \propto (n_{xz} - n_{yz})$. This point of view is supported by recent angle-resolved photoemission~\cite{Yi-ARPES} and nuclear quadrupolar resonance (NQR) measurements~\cite{NQR-1111} and corroborated by \emph{ab initio} calculations~\cite{andriy11}. Alternatively, the electron nematicity may be associated with the $Z_2$-Ising spin ordering~\cite{Fang08, Xu08, Yildirim08} (sometimes referred to as ``spin-nematic''), in which case the order parameter can be written as $\psi \propto (\mathbf{m}_A\cdot\mathbf{m}_B)$  in terms of the sublattice magnetizations $\mathbf{m}_j$ on the square lattice of Fe ions. In both cases, the order parameter $\psi$ is expected to couple linearly to the orthorhombic lattice distortion $\delta$.

While our free energy analysis holds independently of the origin of nematicity, the fact that the two-fold oscillations of the torque magnetization are observable in the strongly overdoped region ($x=0.5$), far away from antiferromagnetic region in the phase diagram (see Fig.\:1 in the main text) suggests that this nematic signal is unlikely of magnetic origin and that orbital physics likely plays an important role.

\end{document}